\documentstyle[
               12pt,
               prd,aps,epsfig]{revtex}

\newenvironment{equation*}{\begin{displaymath}}{\end{displaymath}}

\newcommand{\Scri}{\mbox{$\cal J$}}

\newcommand{\N}[1]{{\cal N}_{\rm\bf #1\>}{}}
\newcommand{\Lien}{{\cal L}_{n}{}}
\newcommand{\DIII}{\,{}^{\scriptscriptstyle(3)\!\!\!\:}\nabla}
\newcommand{\ROI}{\,{}^{\scriptscriptstyle(0,1)\!\!\!\:}\hat R}
\newcommand{\RII}{\,{}^{\scriptscriptstyle(1,1)\!\!\!\:}\hat R}
\newcommand{\epsIII}{\,{}^{\scriptscriptstyle(3)\!\!\!\:}\epsilon}

\newcommand{\hR}{\mbox{$\hat{R}$}}

\newcommand{\tg}{\mbox{$\tilde{g}$}}

\def\@warning#1{\typeout{LaTeX Warning [l.\the\inputlineno]: #1.}}

\begin{document}


\title{How to Avoid Artificial Boundaries in the Numerical Calculation
of Black Hole Spacetimes}

\author{{\bf Peter H\"ubner}\\
        (pth@aei-potsdam.mpg.de)\\
        Max-Planck-Institut f\"ur Gravitationsphysik\\
        Albert-Einstein-Institut\\
        Schlaatzweg 1\\
        D-14473 Potsdam\\
        FRG}

\maketitle
\thispagestyle{empty}
\mbox{}
\\
{\small short title: Calculating Black Hole Spacetimes}
\\
\mbox{}

\begin{abstract}
This is the first of a series of papers describing a numerical
implementation of the conformally rescaled Einstein equation, an
implementation designed to calculate asymptotically flat 
spacetimes, especially spacetimes containing black holes.
\\
In the present paper we derive the new first order time evolution 
equations to be used in the scheme.
These time evolution equations can either be written in symmetric
hyperbolic or in flux-conservative form.
Since the conformally rescaled Einstein equation, also called the
conformal field equations, formally allow us to place the grid boundaries
outside the physical spacetime, we can modify the equations near the
grid boundaries and get a consistent and stable discretisation.
Even if we calculate spacetimes containing black holes, there is no
need for introducing artifical boundaries in the physical spacetime,
which then would complicate, influence, or even exclude the computation of
certain spacetime regions.
\end{abstract}
%
%
%

%
%
\section{Introduction}
Since the capacity of computers has been rapidly growing over the
last years, numerical experiments on spacetimes without symmetries are
now in the reach of the available hardware resources. 
Numerical relativity is on the way to take the step which has already
been taken in other fields of physics: Numerical experiments
supplement, or sometimes even replace, real experiments.
\\
When designing the computer programs for the numerical experiments it
is highly important to be aware of the following significant
difference to other fields of physics:
One of the major methods in developing codes for numerical experiments
has been the comparison and testing with real experiments. 
In numerical relativity however, such possibilities are very rare and
limited; appropriate experiments are not available, comparison with
observations of astrophysical events requires the knowledge of many
unknown parameters, and the known exact solutions presumably do not
represent the full spectrum of properties of solutions of the
non-linear Einstein equation.
\\
The author, therefore, believes, that reliability of the results of
numerical experiments in relativity can only be ensured by
mathematical rigor of the implemented algorithms, i.~e.\ their
vanishing grid size limit should be consistent with the Einstein
equation.

In this series of papers we describe a numerical implementation of the
conformal field equations~\cite{Fr81ot,Fr81ta}. 
In this implementation we do not use any approximation besides the
idealisation of isolated systems and the replacement of the partial
differential equations by their discretised versions.
The approach described below is designed to be able to calculate
global as well as local properties of spacetimes arising from data of
arbitrary strength.
Important issues are e.~g.\ testing the validity range of
approximations, calculating the long-time behaviour of fields,
and investigating the nature of singularities.
For spherically symmetric models with scalar fields the attainability
of these goals has been demonstrated in \cite{Hu93nu,Hu96mf}.
\\
The approach described is so powerful that we can use the same time
evolution scheme to calculate scenarios like the decay of weak
gravitational waves to Minkowski space or the merging of black holes.
\\
The use of the conformal field equations allows us to extend the
calculation to null infinity. 
Moreover, we can formally even extend the spacetime through null
infinity and place the grid boundaries in a region which is causally
disconnected from the physical spacetime.
Then boundary effects, like a violation of the constraints or creation
of spurious backscattering, cannot propagate into the physical
spacetime.
As null infinity is part of the grid, we can calculate anything which
is suggested by a Penrose diagram, including quantities which are only 
defined in the limit of going to infinity, like gravitational
radiation.

Since an exhaustive description of the new ingredients needed for this
(n+1)-dimensional (n$\le$3) implementation of the conformal approach
is too long for a single paper and requires techniques from
mathematical relativity as well as numerical mathematics, we have
split it into a series of papers.
In this first part of the series we describe the foundation of the
time evolution part of the code, namely the geometrical background,
the equations used, and the physical scenarios which will be treated
in detail in the following articles.
In particular, we develop the mathematical basis and give a preview
over the result which can be expected from our numerical approach.
\\
Details about the initial data problem, the discretisation used, the
tests, and more will be found in the following parts of the series. 

In section~\ref{KonfForm} the conformal field equations are written in
first order form and split into symmetric hyperbolic time evolution
equations and constraints.
When deriving this new form, which could also be written in flux
conservative form, we proceed in analogy to~\cite{Fr96hr}, a paper
dealing with the Einstein equations for vacuum in physical spacetime.
Also, these properties of the conformal field equations are recalled
which are essential for the following.
Having the conventions available we discuss the hyperboloidal initial
value problem in section~\ref{AWP}.
There we also explain the special role of the conformal factor,
special not as a variable of the system, but special when the physical
spacetime is reconstructed from the variables of the conformal field
equation.
Section~\ref{Randbehandlung} describes possible grid boundary
treatments and proves their numerical stability.
\\
In the last section we outline the power of the conformal approach
by describing why we can in principle calculate the complete future
evolution of the initial data, including null infinity.
Since conformal rescaling preserves the null cone structure of a
spacetime, the conformal approach is especially tuned for 
analysing causal structure and gravitational radiation.
%
%
%

%
%
\section{A symmetric hyperbolic first order system for the conformal
  field equations}
\label{KonfForm}
We start the derivation of the first order equations by giving the
conformal field equations in our notation and recalling some of their
properties, whose proofs can be found in~\cite{Fr91ot,Hu93nu}.
\subsection{The conformal field equations and the reconstruction of
  physical spacetime}
\label{BasisGl}
In the conformal approach we solve the initial value problem for the
conformal field equations, which read
\begin{mathletters}
\label{KonfGl}
\begin{eqnarray}
\label{Riccigleichung}
\nabla_a \hR_{bc} - \nabla_b \hR_{ac}
+ \frac{1}{12} \left( (\nabla_a R) \, g_{bc} - (\nabla_b R) \, g_{ac} \right)
+ 2 \, (\nabla_d \Omega) \, d_{abc}{}^d & = & 0,
\\
\label{Weylgleichung}
\nabla_d d_{abc}{}^d & = & 0,
\\
\label{OmGl}
\nabla_a \nabla_b \Omega_a + \frac{1}{2} \, \hR_{ab} \, \Omega 
- \frac{1}{4} \nabla^a\nabla_a\Omega \, g_{ab} & = & 0,
\\
\label{omWllngl}
\frac{1}{4} \nabla_a \left(\nabla^b\nabla_b \Omega\right) + 
\frac{1}{2} \, \hR_{ab} \, \nabla^b \Omega 
+ \frac{1}{24} \, \Omega \, \nabla_a R + \frac{1}{12} \, \nabla_a
\Omega \, R 
& = & 0,
\\
\label{irrRiemann}
  \Omega d_{abc}{}^d
  + ( g_{ca} \hR_b{}^d - g_{cb} \hR_a{}^d 
  - g^d{}_a \hR_{bc} + g^d{}_b \hR_{ac} )/2
  + ( g_{ca} g_b{}^d - g_{cb} g_a{}^d ) \frac{R}{12} & = &
  R_{abc}{}^d,
\\
\noalign{and}
\label{Rtrace}
  \Omega^2 R + 6 \, \Omega \, \nabla^a \nabla_a \Omega -
    12 \, (\nabla^a \Omega) \, (\nabla_a \Omega) & = & 0.
\end{eqnarray}
\end{mathletters}
Here the tensor $g_{ab}$ is a Lorentzian metric on a manifold $M$,
$\hR_{ab}$ a symmetric-traceless tensor field, $d_{abc}{}^d$ a tensor
field with the symmetries of the Weyl tensor, $\Omega$ a scalar,
$R_{abc}{}^d$ the differential expression of the Riemann tensor in
terms of the metric~$g_{ab}$, $R$ a prescribed function on $M$, and
$\nabla_a$ denotes the derivative operator associated to~$g_{ab}$. 
\\
For any solution $(g_{ab},\hR_{ab},d_{abc}{}^d,\Omega)$ of
system~(\ref{KonfGl}) the function $R$ turns out to be the Ricci
scalar, $\hR_{ab}$ the traceless part of the Ricci tensor, and $\Omega
d_{abc}{}^d$ the Weyl tensor of~$g_{ab}$.
\\
Any solution of~(\ref{Riccigleichung}--\ref{omWllngl}) solves
(\ref{Rtrace}) on a connected domain if~(\ref{Rtrace}) holds at one
point in the domain.
It is, therefore, sufficient to enforce (\ref{Rtrace}) as a constraint
on the initial data. 
In the following this requirement will no longer be explicitly
mentioned.
\\
We will arrange our calculation such that $(M,g_{ab})$ will be a
globally hyperbolic manifold sliced by the spacelike hypersurfaces
$\Sigma_t$.
By solving~(\ref{KonfGl}) on $M$ we obtain a solution of the
Einstein equation on the physical spacetime $\tilde{M}:= \{x\in
M\>|\>\Omega(x)>0\}$, since for any sufficiently smooth solution
of~(\ref{Riccigleichung}--\ref{omWllngl}) the metric
$\tg_{ab}:=\Omega^{-2} g_{ab}$ is a solution of the Einstein equation
on $\tilde{M}$.
\\
It is sufficient for our purposes to only consider setups for which
the physical part $\tilde \Sigma_{t_0} := \{ x\in \Sigma_{t_0}
\>|\>\Omega(x)>0\}$ of the initial hypersurface $\Sigma_{t_0}$ is
relatively compact.
We define ${\cal S}_t := \{ x\in \Sigma_t\>|\>\Omega(x)=0\}$. 
\\
The part of the boundary of $\tilde{M}$ with $\nabla_a \Omega
\mid_{\Omega=0} \neq 0$ is denoted $\Scri{}$. 
All geodesics of $\tg_{ab}$ which end at \Scri{} are null geodesics. 
Since they intersect \Scri{} at infinite affine parameter with respect
to $\tg_{ab}$, \Scri{} is infinitely far away with respect to $\tg_{ab}$.
\Scri{} represents, therefore, null infinity, and quantities like
gravitational radiation, which is extracted most effectively by taking 
the limit of going to infinity, can be determined by reading off the
values of certain variables at \Scri{}.
\\
As can be seen immediately from equation~(\ref{Rtrace}), \Scri{} is
a null hypersurface.
When solving the initial value problem for data given on the spacelike
hypersurface $\Sigma_{t_0}$, we will, unless mentioned otherwise,
require $\nabla^a\Omega$ to be future directed everywhere on ${\cal S}
:= \{ P\in \Sigma_{t_0} \mid \Omega(P)=0 \}$.
Or stated more interpretatively, we choose the initial hypersurface in
such a way that it intersects future null infinity and not past null
infinity. 
Then $\tilde\Sigma_{t_0} := \tilde{M}\cap\Sigma_{t_0}$ is a Cauchy surface for the
part of $\tilde M$ which lies in the future of $\Sigma_{t_0}$, i.~e.\
$\tilde{M}\cap {\cal D}(\Sigma_{t_0}) = {\cal   D}(\tilde{M}\cap\Sigma_{t_0})$,
where ${\cal D}$ denotes the future domain of
dependence.\footnote{${\cal D}(S) := \{P\in M \mid \mbox{Every past
    inextendible causal curve through $P$ intersects $S$}\}$}
This property guarantees that at the points of $\{x|\>\Omega(x)\ge0\}$ 
the solution of the initial value problem is independent of the
values of the variables at the points in $\{x|\>\Omega(x)<0\}$.
Hence, by placing the grid boundaries outside physical spacetime,
namely into the set $\{x|\>\Omega(x)<0\}$, we gain a decoupling of the
boundary treatment from the calculation of physical spacetime.
\\
If $\Scri{}=R\times{\cal S}$ consists of (disconnected) pieces with
topology $R\times S^2$, the spacetime is called (weakly) asymptotically
flat~\cite{Pe67so}.
The gravitational wave scenarios usually treated fall into this class.
The conformal field equations also allow us to study data with other
topologies of~${\cal S}$, and in fact many aspects of gravitational
radiation can also be studied on such spacetimes.
We will see in subsection~\ref{ToroidalScris}, that in numerical
studies spacetimes with toroidal sections of \Scri{} may provide
advantages over those with $\Scri{}=R\times S^2$, especially in the
cases where we reduce the space dimensions by assuming symmetries.
To have a term available which refers to both cases we call
spacetimes {\bf asymptotically regular},\footnote{regular
  with respect to the existence of null infinity, the spacetime may
  nevertheless contain singularities.} if they arise from regular data for the
conformal field equations with $\nabla_a\Omega$ future directed
everywhere on ${\cal S}_{t_0}$. 
\\
The conformal and the physical metric are related by a rescaling
which is essentially arbritrary, as two solutions $(M,g_{ab},\Omega)$ and
$(M,\bar{g}_{ab},\bar{\Omega})$ with $(\bar{g}_{ab},\bar{\Omega}) =
(\theta^2 g_{ab},\theta \Omega)$ and a positive function $\theta$
describe the same physical spacetime.
Under the rescaling $\theta$ the Ricci scalar $R$ changes.
To obtain equations with nice properties we do not prescribe $\Omega$
but $R$.
The prescribed function $R$ is therefore a gauge source function for
the rescaling gauge.
\\
The equations~(\ref{Riccigleichung}), (\ref{Weylgleichung}) and
(\ref{omWllngl}) are third order equations for the metric $g_{ab}$ and
the conformal factor $\Omega$.
To be able to apply standard theorems which imply well-posedness of the
initial value problem, we introduce appropriate gauge conditions and
formulate the equations~(\ref{KonfGl}) as a symmetric hyperbolic first
order system.
There are various possibilities to do this.
In all variants gauge source functions must be specified to
make the solution of the initial value problem unique.
In the variants which use the spinor~\cite{Fr91ot,Fr97nta} or the
frame formalism~\cite{Fr83cp,Hu95gr} the coordinate gauge freedom is
fixed by 10~gauge source functions (4~degrees of freedom for the
coordinates, 6~for the tetrad).
In the variant to be described here, which uses a 3-tensor formalism,
the coordinate gauge freedom is fixed by the three components of the
shift vector and a function which is closely related to the lapse
function.
\\
The possible implementations of the conformal field equations differ
slightly in the number of variables, the gauge conditions, and the
type of non-linearity.
Our propagation equations allow us to write them in flux conservativ
form.
Whether one of the implementations has significant advantages over the others in
numerical calculations remains to be seen. 
A numerical comparison with~\cite{Fr97nta,Fr97ntb} is planned (see
also subsection~\ref{ToroidalScris}).
\\
Many quantities of physical interest, like gravitational radiation
or tidal forces (geodesic deviation), can be expressed in terms of the
components of the conformal Weyl tensor $d_{abc}{}^d$ or the physical
Riemann tensor.
As the systems just mentioned contain the components of the conformal
curvature as variables, those quantities can be calculated by pure
algebra and their error can be estimated by a convergence analysis.
\subsection{The 3+1 split of the conformal field equations}
\subsubsection{Equations for the lower order quantities}
Proceeding in analogy to~\cite{Fr96hr} we introduce the extrinsic
curvature $k_{ab}$ as new variable.
It is related to the 3-metric\footnote{definition follows}
$h_{ab}$ and the hypersurface
normal\addtocounter{footnote}{-1}\footnotemark{} $n^a$ by the equation 
\begin{mathletters}
\label{4Gleichungen}
\begin{eqnarray}
\Lien h_{ab} - 2 k_{ab} & = & 0.
\end{eqnarray}
The connection $\Gamma^a{}_{bc}$ relates to the curvature
  $R_{abc}{}^d$ by
\begin{eqnarray}
- \partial_a \Gamma^d{}_{bc} + \partial_b \Gamma^d{}_{ac} 
+ \Gamma^e{}_{ca} \Gamma^d{}_{be} - \Gamma^e{}_{cb} \Gamma^d{}_{ae}
- R_{abc}{}^d & = & 0.
\end{eqnarray}
Equation~(\ref{OmGl}) and (\ref{omWllngl}), the
equations for the conformal factor $\Omega$, are written in first
order form as follows,
\begin{eqnarray}
\nabla_a \Omega - {}^{\scriptscriptstyle(4)\!\!\!\:}\Omega_a & = & 0,
\\
\nabla_a {}^{\scriptscriptstyle(4)\!\!\!\:}\Omega_b + 
\frac{1}{2} \, \hR_{ab} \, \Omega 
- \omega \, g_{ab} & = & 0,
\\
\noalign{with
  $\omega=\frac{1}{4}\nabla^a\nabla_a\Omega$, and}
\nabla_a \omega + \frac{1}{2} \, \hR_{ab} \,
{}^{\scriptscriptstyle(4)\!\!\!\:}\Omega^b
+ \frac{1}{24} \, \Omega \, \nabla_a R + 
\frac{1}{12} \, {}^{\scriptscriptstyle(4)\!\!\!\:}\Omega_a \, R & = & 0.
\end{eqnarray}
In the equation for the tracefree Ricci tensor $\hR_{ab}$ we
substitute the derivatives of $\Omega$,
\begin{eqnarray}
\nabla_{[a} \hR_{b]c} 
+ \frac{1}{12} (\nabla_{[a} R) \, g_{b]c} 
+ {}^{\scriptscriptstyle(4)\!\!\!\:}\Omega_d \, d_{abc}{}^d & = & 0. 
\end{eqnarray}
The equation for the conformal Weyl tensor $d_{abc}{}^d$,
\begin{eqnarray}
  \nabla_d d_{abc}{}^d & = & 0,
\end{eqnarray}
is kept. 
\end{mathletters}
\\
We now make a 3+1 split of the system~(\ref{4Gleichungen}) by a
straightforward extension of the treatment given in~\cite{Fr96hr}.
For completeness the essential definitions are repeated but for
further clarification the reader is advised to consult this
reference.
\\
The globally hyperbolic manifold $(M,g_{ab})$ is sliced with spacelike
hypersurfaces $\Sigma_t$. 
The parameter $t$ then provides a natural time coordinate and we can
introduce a timelike vector field $t^a$ related to $t$ by $t^a
\nabla_a t = 1$.
Denoting the unit normal of the hypersurface $\Sigma_t$ by $n^a$,
$n^a$ and $t^a$ are related by
\begin{equation}
  n^a = \frac{1}{N} ( t^a - N^a ),
\end{equation}
with $N$ the lapse function and $N^a$ the shift vector.
The metric $g_{ab}$ induces a 3-metric $h_{ab}$ on each
$\Sigma_t$ by
\begin{equation}
  g_{ab} = h_{ab} - n_a n_b.
\end{equation}
We define 
\begin{equation}
  a_a := n^b \nabla_b n_a.
\end{equation}
The object $a_a = \frac{h_a{}^b \partial_b N}{N}$ is purely spatial.%
\footnote{can be calculated on $\Sigma_t$ by only knowing the values of $N$,
$N^a$, and $h_a{}^b$ on $\Sigma_t$}
Later the coordinate gauge freedom is fixed by specifying the three
components of the shift vector $N^a$ and the function $q :=
\ln(\frac{N}{\sqrt{h}})$ as free functions of space and time.
The following holds:
\begin{equation}
  a_a = h_{a}{}^b \partial_b q + \gamma^b{}_{ab}. 
\end{equation}
$\gamma^a{}_{bc}$ is the 3-connection for the 3-metric $h_{ab}$
and is used as a variable of the first order system.
\\
The 4-connection $\Gamma^a{}_{bc}$ is decomposed into
3-parts by means of the hypersurface normal, 
\begin{eqnarray}
  \Gamma^a{}_{bc} & = &
     {}^{\scriptscriptstyle(1,1,1)\!\!\!\:}\Gamma^a{}_{bc}
     - n^a \>{}^{\scriptscriptstyle(0,1,1)\!\!\!\:}\Gamma_{bc}
     - {}^{\scriptscriptstyle(1,0,1)\!\!\!\:}\Gamma^a{}_{c}\> n_b
     - {}^{\scriptscriptstyle(1,0,1)\!\!\!\:}\Gamma^a{}_{b}\> n_c
\nonumber \\
& &
{}   + {}^{\scriptscriptstyle(1,0,0)\!\!\!\:}\Gamma^a \> n_b n_c
     + n^a \>{}^{\scriptscriptstyle(0,0,1)\!\!\!\:}\Gamma_b \> n_c
     + n^a \>{}^{\scriptscriptstyle(0,0,1)\!\!\!\:}\Gamma_c \> n_b
     - n^a \>{}^{\scriptscriptstyle(0,0,0)\!\!\!\:}\Gamma \> n_b n_c,
\end{eqnarray}
where a ${}^{(\ldots,0,\ldots)}$ in front of a quantity indicates contraction
with $n^a$ and a ${}^{(\ldots,1,\ldots)}$ means projected with $h_a{}^b$,
e.g. ${}^{\scriptscriptstyle(1,0,1)\!\!\!\:}\Gamma^a{}_{c} := 
h^a{}_d n^b h_c{}^e \Gamma^d{}_{be}$. A calculation shows
\begin{mathletters}
\begin{eqnarray}
  && \>{}^{\scriptscriptstyle(0,0,0)\!\!\!\:}\Gamma\> =
    - \frac{n^a \partial_a N}{N}
\\
  && \>{}^{\scriptscriptstyle(1,0,0)\!\!\!\:}\Gamma^a\> =
    \frac{n^b \partial_b N^a}{N} + a^a
\\
  && \>{}^{\scriptscriptstyle(0,0,1)\!\!\!\:}\Gamma_b\> =
    - a_b
\\
  && \>{}^{\scriptscriptstyle(0,1,1)\!\!\!\:}\Gamma_{bc}\> =
    - k_{bc}
\\
  && \>{}^{\scriptscriptstyle(1,0,1)\!\!\!\:}\Gamma^a{}_{b}\> =
    k^a{}_b + \frac{1}{N} h_b{}^c \partial_c N^a
\\
  && \>{}^{\scriptscriptstyle(1,1,1)\!\!\!\:}\Gamma^a{}_{bc}\> =
    \gamma^a{}_{bc}
\end{eqnarray}
\end{mathletters}
For the decomposition of the curvature variables we get
\begin{eqnarray}
  \hR_{ab} & = & 
    \RII_{ab} - n_a \ROI_b - \ROI_a n_b + 
    n_a n_b \>{}^{\scriptscriptstyle(0,0)\!\!\!\:}\hat R\>
\\
\noalign{and}
  d_{abcd} & = & 
    l_{db} E_{ac} - l_{da} E_{bc}
    - l_{cb} E_{ab} + l_{ca} E_{bd}
\nonumber \\
& &
{}  - n_b B_{ae} \epsIII^e{}_{dc}
    + n_a B_{bf} \epsIII^e{}_{dc}
    - n_d B_{cf} \epsIII^e{}_{ba}
    + n_c B_{df} \epsIII^e{}_{ba},
\end{eqnarray}
with $l_{ab} = h_{ab} + n_a n_b$, $\epsIII_{bcd} = n^a
\epsilon_{abcd}$, $E_a{}^a=0$, and $B_a{}^a=0$. As $\hR_{ab}$ is tracefree, 
$\>{}^{\scriptscriptstyle(0,0)\!\!\!\:}\hat R\> = \RII_a{}^a$.
\\
For the derivatives of $\Omega$ we substitute
\begin{eqnarray}
  & & \Omega_0 := n^a \nabla_a \Omega = 
  n^a \, {}^{\scriptscriptstyle(4)\!\!\!\:}\Omega_a\\
\noalign{and}
  & & \Omega_a := h_a{}^b \nabla_b \Omega =
  h_a{}^b {}^{\scriptscriptstyle(4)\!\!\!\:}\Omega_b.
\end{eqnarray}
The set of 3-variables is now complete, it is $(h_{ab}$,
$k_{ab}$, $\gamma^a{}_{bc}$, $\ROI_a$, $\RII_{ab}$, $E_{ab}$,
$B_{ab}$, $\Omega$, $\Omega_0$, $\Omega_a$, $\omega$).
\subsubsection{The split into time evolution equations and constraints}
Contracting all indices of the equations~(\ref{4Gleichungen})
with $n^a$ respectively $h_a{}^b$ and using the new variables
we obtain time evolution equations (equations with $\partial_t$)
and constraints (equations without $\partial_t$).
Although this calculation is lengthy, it is straightforward.
To perform it the author has used a MATHTENSOR program for doing 3+1
decompositions written by himself.
Then, after removing the gauge freedom by fixing gauge source
functions and adding constraints to the time evolution equations
to achieve symmetric hyperbolicity, we obtain a complete set of
symmetric hyperbolic time evolution equations.
We write the equations in terms of the null quantities $\N{}$, the
corresponding equations are obtained by setting the null quantities to
$0$:
\begin{mathletters}
\label{Zeitentwicklungsgleichungen}
\begin{eqnarray}
\label{Nh}
\N{h}_{ab} & = &
{} - \Lien h_{ab} + 2 k_{ab}
\\
\label{Nk}
\N{k}_{ab} & = &
{} - \Lien k_{ab} 
   + \DIII_c \gamma^c{}_{ab}
   + \gamma^{d}{}_{bc} \gamma^c{}_{ad} 
   + a_a a_b + k_{c}{}^{c} k_{ab} 
   - \gamma^{c}{}_{ab} a_c 
\nonumber \\
& &
{} + h_{a}{}^{c} h_{b}{}^{d} \partial_d \partial_c q
- \frac{R}{12} h_{ab} - \RII_{c}{}^{c} h_{ab} - 2 \Omega E_{ab}
\\
\label{Ngamma}
\N{\gamma}^{a}{}_{bc} & = &
{} - h_{ad} \, \Lien \gamma^{d}{}_{bc} 
   + \DIII_a k_{bc}
   - a_a k_{bc} + a_c k_{ab} + a_b k_{ac}
   + h_{da} h_{b}{}^{e} h_{c}{}^{f} \frac{1}{N} \partial_f \partial_e N^d  
\nonumber \\
& &
{} - \frac{1}{2} h_{ac} \ROI_b
   - \frac{1}{2} h_{ab} \ROI_c
   + h_{bc} \ROI_a 
   + \epsIII_{ac}{}^{d} \Omega B_{db}
   + \epsIII_{ab}{}^{d} \Omega B_{dc}
\\
\label{NE}
\N{E}_{ab} & = &
{} - \Lien E_{ab} 
   + \frac{1}{2} \epsIII_{a}{}^{cd} \DIII_d B_{cb}
   + \frac{1}{2} \epsIII_{b}{}^{cd} \DIII_d B_{ca}
   + a^c \epsIII_{cb}{}^{d} B_{da} 
   + a^c \epsIII_{ca}{}^{d} B_{db}
\nonumber \\
& &
{} - h_{ab} k^{cd} E_{cd}
   + \frac{5}{2} k_{b}{}^{c} E_{ca}
   + \frac{5}{2} k_{a}{}^{c} E_{cb}
   - 2 k_c{}^c E_{ab}
\\
\label{NB}
\N{B}_{ab} & = &
{} - \Lien B_{ab} 
   - \frac{1}{2} \epsIII_{a}{}^{cd} \DIII_d E_{cb}
   - \frac{1}{2} \epsIII_{b}{}^{cd} \DIII_d E_{ca}
   + a^c \epsIII_{bc}{}^{d} E_{da} 
   + a^c \epsIII_{ac}{}^{d} E_{db}
\nonumber \\
& &
{} - h_{ab} k^{cd} B_{cd}
   + \frac{5}{2} k_{b}{}^{c} B_{ca}
   + \frac{5}{2} k_{a}{}^{c} B_{cb}
   - 2 k_c{}^c B_{ab} 
\\
\label{NROI}
\N{\ROI}_{a} & = &
{} - \Lien \ROI_{a} + \DIII_b \RII_{a}{}^{b}
   - \frac{1}{4} \DIII_a R - k_{b}{}^{b} \ROI_a 
   + a_b \RII_{a}{}^{b} + a_a \RII_{b}{}^{b}
\\
\label{NRII}
\N{\RII}_{ab} & = &
{} - h_{bc} \, \Lien \RII_{a}{}^{c} 
   + \DIII_a \ROI_{b} - \frac{1}{12} h_{ab} \Lien R
   + a_b \ROI_a + a_a \ROI_b 
\nonumber \\
& &
{} - k_{ab} \RII_{c}{}^{c} - k_{cb} \RII_{a}{}^{c} 
   - 2 \epsIII^{cd}{}_b \Omega_c B_{da} - 2 \Omega_0 E_{ab}
\\
\label{NOm}
\N{\Omega} & = &
{} - \Lien \Omega + \Omega_0
\\
\label{NOmO}
\N{\Omega_0} & = &
{} - \Lien \Omega_0 - \omega + a^a \Omega_a - \frac{\Omega}{2} \RII_{a}{}^{a}
\\
\label{NOmI}
\N{\Omega_a}_a & = &
{} - \Lien \Omega_a + a_a \Omega_0 + k_{a}{}^{b} \Omega_b
   - \frac{\Omega}{2} \ROI_a
\\
\label{Nom}
\N{\omega} & = &
{} - \Lien \omega - \frac{\Omega}{24} \Lien R 
   - \frac{R}{12} \Omega_0 - \frac{\Omega^a}{2} \ROI_a 
   + \frac{\Omega_0}{2} \RII_{a}{}^{a}
\end{eqnarray}
\end{mathletters}
Note that for a 3-tensor $t_{\cdots a\cdots}{}^{\cdots b\cdots}$ the following
equality holds:
\begin{equation*}
  \Lien t_{\cdots a\cdots}{}^{\cdots b\cdots} =
     \frac{1}{N} \left( {\cal L}_{t} - {\cal L}_{N} \right)
     t_{\cdots a\cdots}{}^{\cdots b\cdots} + 
     \ldots + n^b \, t_{\cdots a\cdots}{}^{\cdots c\cdots} \, a_c + \ldots
\end{equation*}
The operator $\DIII_a$ is the covariant derivative preserving the 3-metric
$h_{ab}$.
\\ 
If one uses new variables for $E_{ab}$ and $B_{ab}$ as
in~\cite{Fr96hr}, this system is indeed symmetric hyperbolic.
\\ 
The equations~(\ref{Nh}--\ref{NB}) have the same principal part as the
corresponding equations~(6.4--6.8) in~\cite{Fr96hr}. The principal
parts of (\ref{Nh}) and (\ref{NOm}--\ref{Nom}) are equal, the
same is true for the principal parts of the
subsystems~(\ref{Nk},\ref{Ngamma}) and
(\ref{NROI},\ref{NRII}). 
Therefore the system~(\ref{Zeitentwicklungsgleichungen}) possesses the same
characteristics as the equations~(6.4--6.8) in~\cite{Fr96hr}, namely they
are either null hypersurfaces with respect to $g_{ab}$ or they are
timelike and tangent to $n^a$ or to the timelike cone $\{c_{ab} t^a t^b =
0\}$ where $c_{ab} = 4 h_{ab} - n_a n_b$. 
Since all characteristics are within the light cone of $g_{ab}$, the
system is consistent with Einstein causality.
\\
The treatment of the coordinate gauge freedom (fixed by
the functions $q$ and $N^a$) is also discussed in more detail
in~\cite{Fr96hr}. 
What is said there extends straightforwardly to the rescaling freedom
fixed by the gauge source function $R$. 
\\
For the numerical treatment the equation~(\ref{Ngamma}) is multiplied
by $h^{ea}$ and the $h_{bc}$ in (\ref{NRII}) is moved into the Lie
derivative. Both manipulations are done to get $\frac{1}{N}$ times
the identity matrix in front of the time derivative and the second is
also done to obtain the 6 components of the symmetric tensor
$\RII_{ab}$ instead of the 9 components of $\RII_{a}{}^{b}$ 
as variables of the system.
The two equations resulting from
$\N{\RII_{\underline{a}\:\!\underline{b}}}$ and
$\N{\RII_{\underline{b}\:\!\underline{a}}}$ are 
identical only up to constraints and we take the arithmetic mean of 
the two null quantities.
By this averaging some characteristics change but they remain within
the light cone of $g_{ab}$.
\\
The system~(\ref{Zeitentwicklungsgleichungen}) can also be written in
flux conservative form. 
The results we obtained by solving the flux
conservative system with a rotated Richtmyer scheme were less accurate
than those obtained by solving the quasilinear
form~(\ref{Zeitentwicklungsgleichungen}) with the same scheme.
Furthermore calculations on a simple non-linear model system showed
that damping of grid modes is weaker for the flux-conservative form.
Therefore, the quasilinear form of the equations seems to be better
suited for our application than the flux conservative form and we 
refrain from presenting the flux conservative form of the system.
\\[\baselineskip]
The null quantities not appearing in the
system~(\ref{Zeitentwicklungsgleichungen}) are the constraints:
\begin{mathletters}
\label{Constraints}
\begin{eqnarray}
\N{h}_{abc} & = & \DIII_a h_{bc}
\\
\N{k}_{abc} & = &
{} - \DIII_a k_{bc} + \DIII_b k_{ac}
   + \frac{1}{2} h_{ca} \ROI_b - \frac{1}{2} h_{cb} \ROI_a
   - \epsIII_{ab}{}^d \Omega B_{dc}
\\
\N{\gamma}_{abc}{}^d & = & 
{} - \DIII_a \gamma^d{}_{bc}
   + \DIII_b \gamma^d{}_{ac}
   + \gamma^d{}_{ae} \gamma^e{}_{bc}
   - \gamma^d{}_{be} \gamma^e{}_{ac}
\nonumber \\
& &
{} - k_a{}^d k_{bc}
   + k_{ac} k_b{}^d
   + \frac{1}{12} h_a{}^d h_{bc} R
   - \frac{1}{12} h_{ac} h_b{}^d R
\nonumber \\
& &
{} - \frac{1}{2} h_b{}^d \RII_{ac} 
   + \frac{1}{2} h_a{}^d \RII_{bc}
   - \frac{1}{2} h_{ac} \RII_{b}{}^d
   + \frac{1}{2} h_{a}{}^d \RII_{bc}
\nonumber \\
& &
{} - h_{ac} \Omega E_b{}^d 
   + h_{bc} \Omega E_a{}^d
   + h_b{}^d \Omega E_{ac}
   - h_a{}^d \Omega E_{bc}
\\
\N{E}_a & = & 
{} - \DIII_b E_a{}^b
   - \epsIII_{abc} k^{bd} B_d{}^c
\\
\N{B}_a & = & 
{} - \DIII_b B_a{}^b
   + \epsIII_{abc} k^{bd} E_d{}^c
\\
\N{\ROI}_{ab} & = & 
{} \DIII_a \ROI_b - \DIII_b \ROI_a
   + k_b{}^c \RII_{ca} - k_a{}^c \RII_{cb}
   + 2 \epsIII_{ab}{}^c \Omega_d B_c{}^d
\\
\N{\RII}_{abc} & = & 
{} \DIII_a \RII_{bc} - \DIII_b \RII_{ac}
   - \frac{1}{12} h_{ac} \DIII_b R + \frac{1}{12} h_{bc} \DIII_a R
   + \ROI_a k_{bc} 
\nonumber \\
& &
{} - \ROI_b k_{ac} 
   + 2 \epsIII_{ab}{}^d \Omega_0 B_{dc}
   - 2 \Omega_a E_{bc} + 2 \Omega_b E_{ac} 
   + 2 h_{ca} \Omega_d E_{b}{}^d - 2 h_{cb} \Omega_d E_{a}{}^d
\\
\N{\Omega}_{a} & = & {} - \DIII_a \Omega + \Omega_a
\\
\N{\Omega_0}_{a} & = &
{} - \DIII_a \Omega_0 + k_a{}^b \Omega_b - \frac{1}{2} \Omega \ROI_a
\\
\N{\Omega_a}_{ab} & = &
{} - \DIII_a \Omega_b + h_{ab} \omega + k_{ab} \Omega_0
   - \frac{1}{2} \Omega \RII_{ab}
\\
\N{\omega}_{a} & = &
{} - \DIII_a \omega 
   - \frac{1}{24} \Omega \DIII_a R
   - \frac{1}{12} \Omega_a R
   + \frac{1}{2} \Omega_0 \ROI_a
   - \frac{1}{2} \Omega^b \RII_{ba}
\end{eqnarray}
\end{mathletters}
The constraints are, just as the time evolution equations, regular
on the whole manifold $M$ including \Scri{}.
We abstain from proving the propagation of the constraints under the
time evolution equations for the first order form of the conformal
field equations derived.
Since the propagation of the constraints is supported by the following
circumstances, there is no serious doubt that they do propagate.
First, the propagation has been proven for the forms of the conformal
field equations given in \cite{Fr85ot,Hu95gr,Fr97nta}.
Second, for the first order form of the Einstein equation derived
in~\cite{Fr96hr}, which is obtained from the form given by setting
$R=0$, $\ROI_a=0$, $\RII_{ab}=0$, $\Omega=1$ and hence $d_{abc}{}^d =
\tilde C_{abc}{}^d$, the propagation of the constraints has been
proven.
And, third, under the numerical time evolution the constraints propagate
modulo the order of convergence of the scheme used.
\\
To give initial data we have to find a solution of the constraints.
With the exception of cases with high symmetry~\cite{Hu93nu}, 
it is not known yet how to solve the constraints~(\ref{Constraints})
directly for the variables $\underline{f}:=(h_{ab}, k_{ab},
\gamma^a{}_{bc}, E_{ab}, B_{ab}, \ROI_a, \RII_{ab}, \Omega, \Omega_0,
\Omega_a, \omega)$.
Nevertheless the existence of solutions of the constraints on
$\tilde\Sigma\cap\partial\tilde\Sigma$ has been proven
in~\cite{AnCA92ot} for arbitrary topology of $\partial\tilde\Sigma$.
After solving an elliptic problem for
$(h_{ab},k_{ab},\Omega,\Omega_0)$, appropriate contractions
of~(\ref{Constraints}) can be used to calculate the remaining data.
It is shown in~\cite{AnCA92ot}, that, although divisions by $\Omega$
are needed to calculate the curvature related quantities, these
curvature quantities are smooth at $\partial\tilde\Sigma$, where $\Omega$
vanishes ($={\cal S}$), and provide, therefore, suitable data for the
time evolution equations.
Clearly, this division by $\Omega$ is numerically a delicate issue.
However, in~\cite{HuInPrep} we will discuss a way of doing this which
works for arbitrary ${\cal S}$.
%
%

%
%
\section{The role of the conformal factor}
\label{AWP}
In the preceeding section we have derived symmetric hyperbolic
evolution equations of first order representation from the conformal
field equations.
Apart from the treatment of the grid boundaries there are standard
techniques available to obtain stable discretisations of symmetric
hyperbolic systems.
Due to properties distinguishing the conformal approach from
approaches working in terms of physical spacetime the treatment of the outer
grid boundary can be made trivial, as we will see in the next section. 
Therefore, it is interesting to understand what kind of physical
scenarios can be dealt with by the conformal approach without
complicating matters by introducing additional (inner) boundaries.
\\
We will see in this section that the class of scenarios meeting
the requirements just mentioned is much larger than the class we would
get under these requirements when working in terms of physical
spacetime.
The underlying reason is the role of the conformal factor $\Omega$.
For the equations, and therefore for the discretisations, the
conformal factor is just one of many variables which may assume any
value.
But the conformal factor is a very special variable when transforming
back to physical spacetime by rescaling $g_{ab}\mapsto\Omega^{-2}
g_{ab}$.
Then the $\Omega=0$ level sets turn out to be boundaries of the
physical spacetime.
This double role of the conformal factor not only enlarges the class
of physical problems which we can treat by standard techniques of
numerical mathematics, it also determines the character of the initial
value problem we solve.
\subsection{The hyperboloidal initial value problem}
Let $\underline{f}(t_0,\vec x) =
(h_{ab},k_{ab},\gamma^a{}_{bc},E_{ab},B_{ab},\ROI_{a},\RII_{a}{}^{c},
\Omega,\Omega_0,\Omega_a,\omega)$ be a smooth solution of the
constraints on an everywhere spacelike hypersurface
$\bar\Sigma_{t_0}$, $\Omega\mid_{\bar\Sigma_{t_0}}\ge 0$, 
with boundary ${\cal S}_{t_0}$, where $\Omega$ vanishes.
If $\nabla^a \Omega\mid_{\cal S}$ is a future directed, non-vanishing
null vector, we call $(\bar\Sigma_{t_0},\underline{f}(t_0,\vec x))$ a
{\bf hyperboloidal initial value problem}.
\\
We can smoothly extend the hypersurface $\bar\Sigma_{t_0}$ and the
data $\underline{f}(t_0,\vec x)$ beyond the boundary to obtain an
{\bf extended hyperboloidal initial value problem}
$(\Sigma_{t_0},\underline{f}(t_0,\vec x))$.
Without loss of generality we assume that $\Omega<0$ on the extension.
For an extended hyperboloidal initial value problem the physical part
of the future of $\Sigma_{t_0}$ is independent of the formal extension
of the initial data beyond ${\cal S}$ since $\tilde M\cap{\cal
  D}(\Sigma) = {\cal D}(\tilde M\cap\Sigma)$.
Therefore, we do not require the extended data to satisfy the
constraints outside $\bar\Sigma_{t_0}$.
\subsection{Interpretation of various topologies of ${\cal S}$}
\label{Darstellung}
In this subsection we classify various initial data configurations by
finding spacetimes with the same topology of ${\cal S}$.
It may well be that the data in each class develop into spacetimes
with very different structures in the large.
To analyse this variety in detail is an interesting issue and it is a
goal of the numerical experiments.
Here, our purpose is to only describe a minimal collection of
scenarios which we can treat with the conformal approach without
introducing inner boundaries.
\subsubsection{$\tilde\Sigma$s with spherical boundaries}
\label{SphericalScris}
The simplest case of a compact three dimensional manifold ${\cal
  S}\cup\tilde\Sigma$ is shown in figure~\ref{MinkRZ}, namely the ball
$B^3$. 
\begin{figure}[h]
  \begin{center}
      \begin{minipage}[t]{6.4cm}
        \input Minkowski.pstex_t
        \\
        \caption{\label{MinkRZ}Data evolving into a spacetime with
          an asymptotic structure as Minkowski spacetime. 
          The third dimension is suppressed, ${\cal S}$ is a sphere.}
      \end{minipage}
    \hspace{4em}
      \begin{minipage}[t]{6.4cm}
        \input BH.pstex_t
        \\
        \caption{\label{BHRZ}Data evolving into a spacetime with an
          asymptotic structure as Schwarzschild-Kruskal. 
          The third dimension is suppressed. 
          The ${\cal S}$'s are spheres, the corresponding event
          horizons are also shown.}
      \end{minipage}
  \end{center}
\end{figure}%
Its interior is diffeomorphic to $R^3$.
For a standard choice of the metric this is a hyperboloidal slice in
Minkowski spacetime.
For weak data the whole spacetime evolving from the data has an
asymptotic structure similar to Minkowski spacetime~\cite{Fr91ot}. 
Timelike infinity $i^+$ can be represented by a regular point on the
grid.
For strong data the spacetime will develop a singularity, there will
not be any regular $i^+$.
In general, we expect strong field configurations to develop regions
for which all future directed causal curves end in a singularity and
the arising spacetimes will then have at least one event horizon.
\\
The next, slightly more complicated manifold is obtained by cutting
out a ball from the interior of the ball as shown in figure~\ref{BHRZ}.
The interior $\tilde\Sigma$ has topology $R\times S^2$ and there are
two ${\cal S}$'s evolving into two \Scri{}s.
This is the topology of a slice in the conformal picture of the
Schwarzschild-Kruskal spacetime (see also figure~\ref{Kruskal}).
By Birkhoff's theorem spherically symmetric data with this topology of 
${\cal S}$ will evolve into the Schwarzschild-Kruskal spacetime with
its spacelike singularity in the future.
For data close to Schwarzschild data there will probably be a spacelike
singularity in the future and there will then be an event horizon around
each \Scri{}.
Therefore it is justified to speak in the situation of
figure~\ref{BHRZ} of a one black hole spacetime,
although there may be data for which no singularity appears. 
\\
By cutting out $N$ balls from the interior, spacetimes with $N+1$
\Scri{}s are obtained.
Figure~\ref{2BHRZ} shows the $N=2$ case.
\begin{figure}[h]
  \begin{center}
      \begin{minipage}[t]{6.4cm}
        \input 2_BH.pstex_t
        \\
        \caption{\label{2BHRZ}Data evolving into a spacetime with an
          asymptotic structure as a two black hole spacetime
          with three disjoint null infinities. 
          The third dimension is again suppressed. 
          The ${\cal S}$'s are spheres.}
      \end{minipage}
    \hspace{4em}
      \begin{minipage}[t]{6.4cm}
        \input Torus.pstex_t
        \\
        \caption{\label{TorRZ}Data evolving into a spacetime with an
          asymptotic structure as the A3 spacetime.
          The third dimension is again suppressed. 
          The grid boundaries in the vertical dimension are identified
          as well as the boundaries in the suppressed dimension. 
          The ${\cal S}$'s are tori.}
      \end{minipage}
  \end{center}
\end{figure}
Spacetimes usually interpreted as N black holes with N+1 \Scri{}'s
have hyperboloidal hypersurfaces like the one shown.
Whether, in the case $N=2$ for simplicity, an observer at
$\Scri{}_1$ actually sees one or two black holes depends on the slice
and the data.
If the event horizon of $\Scri{}_1$ consists of two pieces, there are
two separated black holes which may merge later, i.~e.\ the two pieces
of the event horizon merge.
\subsubsection{$\tilde\Sigma$s with toroidal and other boundaries}
\label{ToroidalScris}
The examples discussed in the preceeding subsection are all
asymptotically flat spacetimes which are only a subset of the
asymptotically regular spacetimes.
Except for one case spacetimes which are asymptotically regular, but
not asymptotically flat, will not be discussed here.
\\
Although nowadays computers provide the resources for doing three
dimensional calculations with reasonable resolution, high accuracy
requirements as needed for certain interesting scenarios may require
calculations with higher resolution.
Much higher resolutions can be achieved by analysing spacetimes with
one Killing vector and they may also be easier to understand as
they are easier to imagine and to visualise.
In asymptotically flat spacetimes a Killing vector field
with closed orbits vanishes at certain points and therefore there is
an axis.
Coordinate systems adapted to a one Killing vector symmetry become
singular at the axis, the evolution equations or the variables are
singular there, and numerical instabilities are very difficult to avoid.
So it would be very interesting to have spacetimes available which
admit one Killing vector without having an axis, which possess null
infinity, and which contain gravitational radiation.
Those spacetimes are similar to the Schwarzschild solution, which is a 
spherically symmetric solution without a centre.
\\
This was the motivation of B. Schmidt to investigate these
spacetimes~\cite{Sc96vs}.
According to their simplest representative, the A3 solution in the
classification of Ehlers and Kundt~\cite{EhK62es}, they are called 
asymptotically A3. 
Figure~\ref{TorRZ} shows the initial slice for an asymptotically A3
spacetime. 
At the boundary of the suppressed $z$ direction and the vertical $y$
direction points are identified.
Both ${\cal S}$'s are tori.
The A3 solution has a Killing vector with circular
orbits but without axis.
Many other asymptotically A3 solutions with two Killing vectors
are known~\cite{Sc96vs,Hu98ma}.
In contrast to the spherically symmetric Schwarzschild solution
many of them contain gravitational radiation~\cite{Fo97sa}.
\\
Asymptotically A3 solutions provide an excellent testbed for numerical
codes as they allow us the testing of radiation extraction procedures
against exact solutions.
They have been intensely used for testing in the conformal
codes~\cite{Fr97ntb,HuInPrep}.
Details about the tests of the author's code on these examples are
going to be reported in~\cite{HuInPrep}.
\subsection{\Scri{}-fixing}
\label{ScriFixing}
In the hyperboloidal initial value problem the boundaries ${\cal S}$
of $\tilde\Sigma_{t_0}$ are parts of ingoing null hypersurfaces.
For a vanishing shift vector the coordinate area of $\tilde\Sigma_t$ shrinks,
the physical part of the hypersurface is represented by less and less
gridpoints.
This loss of resolution is a wide-spread objection against the numerical
application of conformal techniques. 
But this objection is unjustified as the shift vector can always be chosen
in such a way that the location of ${\cal S}_t$ is fixed and no
gridpoints are lost.
We call such a choice ``\Scri{}-fixing''.
\\
In the spherically symmetric case the author has used such a choice of
coordinates in accuracy tests leading to~\cite{Hu93nu}.
There were two reasons why this choice was not pursued further:
Firstly, the accuracy obtained was lower than the one for the case of
an ``infalling'' \Scri{} even for late times and, secondly, $i^+$ was for
that particular choice of coordinates no longer on the grid, the time
direction was ``decompactified'' [unpublished work].
\\
J.~Frauendiener has generalised the \Scri{}-fixing~\cite{Fr97ntb} to
the non-spherically symmetric case.
In the variables used in this paper the shift vector on \Scri{} has to be
chosen as follows:
\begin{equation}
  \label{ScriFixingOnScri}
  N^a\mid_{\Scri{}} = - \frac{N}{\Omega_0} h^{ab} \Omega_b.
\end{equation}
It is easy to see that~(\ref{ScriFixingOnScri}) plugged into
equation~(\ref{NOm}) yields $\partial_t\Omega|_{\Scri}=0$.
The expression is regular on \Scri{} as $\nabla_a\Omega$ is a
non-vanishing null vector on \Scri{} and therefore $\Omega_0\ne0$, but
$\Omega_0$ may become singular on $M/{\cal S}$.
As an alternative everywhere regular on $M$ one can use:
\begin{equation}
  \label{ScriFixingReg}
  N^a = - N \, \frac{\Omega_0}{(\Omega_0)^2+(\Omega)^2} \, h^{ab} \Omega_b.
\end{equation}
Since $N^a$ depends on variables, equation~(\ref{Ngamma}) contains
second derivatives of variables making the system of
equations~(\ref{Zeitentwicklungsgleichungen}) formally ``parabolic''.
But those derivatives can be eliminated by the use of the other
equations yielding a system first order in space and time again.
J.~Frauendiener has shown that the conformal field equations in spinor
formulation stay symmetric hyperbolic even with
\Scri{}-fixing~\cite{Fr97ntb}.
He also demonstrated that \Scri{}-fixing works in his two dimensional
code.
\\
By using a \Scri{}-fixing choice of shift there is no loss in freedom
of specifying the shift vector on $\tilde M$, as with a \Scri{}-fixing
shift $N^a$ the new shift $N^a + \Omega \tilde N^a$ is also
\Scri{}-fixing for arbritrary $\tilde N^a$.
%
%
%
%

%
%
\section{A way to obtain stable boundaries}
\label{Randbehandlung}
When solving symmetric hyperbolic equations numerically, the discretisation
scheme used must be altered at the boundaries of the grid.
It is not only necessary that the treatment must be consistent with
the initial boundary value problem of the equations, the interior
scheme and the boundary treatment must also fit.
Otherwise gridmodes not related to a solution of the continuous
equation arise from the boundaries and trigger instabilities.
Even for very simple linear equations like the advection equation many
discretisations of the boundary are unstable, although they look
reasonable.
\\
In general relativity the problem is worse, even the analytic
boundary value problem is not completely understood~\cite{FrN98ti}.
On a timelike boundary the evolution equations and the constraints
must be solved simultaneously. 
This consistency requirement determines the number of functions which
can be specified freely.
If the constraints are violated at the boundary, the numbers
calculated at the gridpoints influenced by the boundary will not
approximate any solution of the Einstein equation.
The grid is filled with ``invalid points'' very quickly, independently 
of the grid spacing. 
To make things even worse the error made cannot be estimated by the
size of the violation of the constraints~\cite{Ch91co}.
Therefore a discretisation must not only be stable but also
consistent with constraint propagation.
\\
Even if one could solve this boundary problem there would still remain
the problem of how to suppress artificial backscattering of an
outgoing wave when it hits the boundary. 
Already for the wave equation in two and in three space dimensions that
problem is not solvable by a local procedure~\cite{EnM77ab}, as the
uniquely determined condition for ``no backscattering'' involves a
non-local pseudo-differential operator.
Stable local approximations to this non-local pseudo-differential
operator show typically an artificial reflection of the order of
percent~\cite{EnM77ab}.
\\
If it were necessary to calculate a solution of the conformal
field equations on the whole grid, the situation would be similarly
hopeless.
But we do not have to calculate a solution of the conformal field
equations on the whole grid.
As already pointed out, $\tilde{M}\cap {\cal D}(\Sigma) = {\cal
  D}(\tilde{M}\cap\Sigma)$.
Therefore one can modify the equations and/or the data outside
$\tilde{M}\cup\Scri{}$.
\\
The ``boundary treatment'' we are going to present utilises this
property of the conformal field equations and is independent of the
exact form of the time evolution equations.
To simplify the writing we continue by writing our time evolution
equations in the general form 
\begin{equation}
\label{SystemAbstr}
  \partial_t \underline{f} 
  + \underline{\underline{A}}^i \partial_i \underline{f} - 
  \underline{b} = 0.
\end{equation}
The $\underline{\underline{A}}^i$ are quadratic matrices depending on
$\underline{f}$, $\underline{b}$ is a vector also depending on
$\underline{f}$.
\\
To obtain equations whose boundary treatment has already been analysed 
the equations~(\ref{SystemAbstr}) are modified to 
\begin{equation}
\label{Staubsauger}
  \partial_t \underline{f} + (1-\alpha(\Omega)) \> r^i \partial_i \underline{f}
  + \alpha(\Omega) \left( \underline{\underline{A}}^i \partial_i \underline{f} -
  \underline{b} \right) = 0,
\end{equation}
with a sufficiently often differentiable function $\alpha(\Omega)$,
which is $0$ for $\Omega\leq\Omega_0<\Omega_1<0$ and $1$ for
$\Omega\geq\Omega_1$. 
$r^i$ is an outward pointing vector field.
For $\Omega_0>\max_{\mbox{\scriptsize grid boundary}}\Omega$ one
obtains near the boundary an equation with well-analysed
boundary discretisations for many interior schemes.
For an overview and the proofs of stability see~\cite{AbG79so,AbM81so}.
The local order of convergence at the boundary may be one order lower
than the convergence order $n$ of the interior scheme without
endangering the overall order $n$~\cite{Gu75tc}.
\\
The equations near the boundary are outward directed advection equations.
Numerical noise created by a small transition zone
$[\Omega_0,\Omega_1]$ is transported to the boundary.
Since $\Omega$ does not necessarily stay less than $\Omega_0$, the
stability property of the boundary may get lost.
By choosing $r^i=0$ this can be avoided, the time evolution near the
boundary gets frozen, one obtains:
\begin{equation}
\label{FREEZE}
  \partial_t \underline{f} 
  + \alpha(\Omega) \left( \underline{\underline{A}}^i \partial_i \underline{f} - 
  \underline{b} \right) = 0.
\end{equation}
Of course the possibilities are not restricted to the cases
discussed.
One could e.~g.\ try to combine the advantages of both methods
described by only setting $r^i=0$ in the equation for $\Omega$.
\\
The change of the system~(\ref{SystemAbstr}) to (\ref{Staubsauger})
or (\ref{FREEZE}) will not change the solution on $\tilde M$.
Nevertheless the numerically calculated numbers may change on
$\tilde M$.
In a numerical calculation we have to fulfil the
Courant-Friedrich-Levy condition, hence the numerical range of
influence of a point is larger than or equal to the analytic range of
influence, i.~e.\ values at gridpoints in $\tilde M$ may numerically
depend on values outside \Scri{}.
For a convergent scheme the dependence of the calculated solution on
the values in the difference $\Delta$ of the two ranges
must go to zero with the convergence rate.
The error made by changing the data or the equations in $\Delta$ is of
the same size, order, and nature as the discretisation error.
This error is automatically taken into account when determining the
error bars of the calculated values by a convergence analysis.
Therefore, although the numbers actually calculated as an
approximation to the solution may be changed by the boundary
treatment, the physical predictions made from the numerical
experiments do not change, since the physical predictions of a
numerical experiment are the numbers modulo the error bars.
\\
Beside the simplification of the boundary treatment the change of the
equations outside $\tilde M$ has other positive effects.
The equation~(\ref{SystemAbstr}) may develop singularities everywhere
in the conformal spacetime~\cite{Hu93nu,Hu96mf}.
Since the equations of system~(\ref{Staubsauger}) or (\ref{FREEZE}) cannot
develop singularities for $\Omega<\Omega_0$ the appearance of a
singularity outside $\tilde M$ becomes less likely.
\\
The treatment of the grid boundaries described above is also
independent of the choice of the gauge source functions and the
coordinate representation of the light cone which depends on the gauge
source functions.
For eigenfeld methods, also applicable as boundary
treatment~\cite{Fr97ntb}, this is not the case.
The choice of the gauge source functions at the boundary influences
the number of outward and inward propagating fields and, therefore,
the boundary treatment.
%
%
%

%
%
\section{Aspects of our geometrical setting}
\label{Programm}
In the preceeding sections we have deduced a set of equations,
which is equivalent to the Einstein equation and whose form is adapted
to the requirements of numerics.
To conclude this paper, we will now describe on a heuristic level,
based on the rigorous statements of the preceeding sections, how
we proceed to calculate a spacetime from the initial slice.
We do this on two examples of general interest and compare on those
with alternative approaches.

In the first example we illustrate by means of a Penrose diagram what
part of a spacetime we can in principle calculate. 
To use a sufficiently rich example, i.~e.\ an example for which inner
and outer boundaries were believed to be needed, but still a simple
enough to possess a depictable Penrose diagram, we have chosen the
Schwarzschild-Kruskal spacetime.
\\
In the second example we explain on the scenario of an energy/matter
concentration collapsing to a black hole how gravitational radiation
propagates through the conformal spacetime.
\subsection{Evolving spacetime from an initial slice}
\label{Heuristik}
\begin{figure}[htbp]
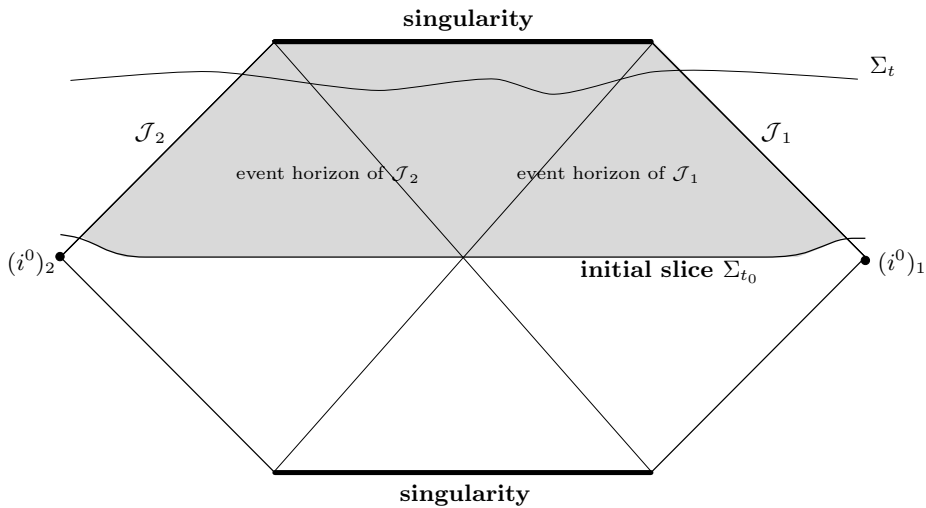

  \begin{center}
      \begin{minipage}[t]{13.2cm}
        \input Kruskal.pstex_t
        \\
        \caption{\label{Kruskal}Penrose diagram of the
          Schwarzschild-Kruskal spacetime showing the region which can 
          be calculated in the conformal approach. All points with
          the exception of the two spacelike infinities $(i^0)_1$ and
          $(i^0)_2$ represent an $S^2$.}
      \end{minipage}
  \end{center}
\end{figure}

In figure~\ref{Kruskal} the Penrose diagram of the Schwarzschild-Kruskal
spacetime is shown.
In the numerical construction of this spacetime from an extended
hyperboloidal initial value problem we give appropriate data on an
extended hyperboloidal slice $\Sigma_{t_0}$.
The grid boundaries are placed outside the region representing the
physical part of the initial slice.
By solving the conformal field equations we calculate a slicing
$\Sigma_{t}$ of the future of $\Sigma_{t_0}$.
\\
In principle there is no reason, assuming appropriate coordinates,
which prevents us from calculating the whole domain of dependence of
the physical part $\tilde\Sigma_{t_0}$ of $\Sigma_{t_0}$ up to the
singularity, i.~e.\ to calculate the solution of the Einstein equation 
in the shaded region of the figure.
The conformal field equations can be viewed as the Einstein equation for
an artificial spacetime with an artificial matter field, namely the
rescaling factor.
With this analogy in mind it becomes obvious that the
relation between coordinate choice and exhaustive slicing of the
spacetime is similar in conformal and physical spacetime.

\begin{figure}[htbp]
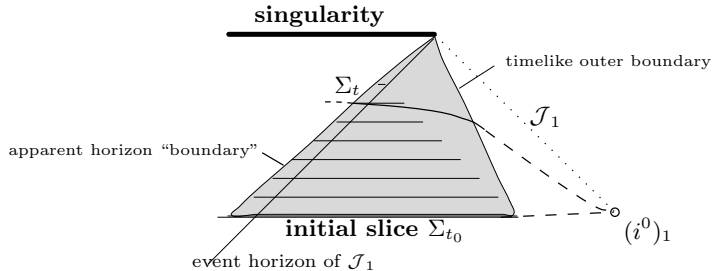

  \begin{center}
      \begin{minipage}[t]{8cm}
        \input CauKruskal.pstex_t
        \\
        \caption{\label{KakKruskal}The part of the Penrose diagram of
          Schwarzschild spacetime which is calculated when solving
          the initial boundary value problem in physical
          spacetime. Again every point represents an $S^2$.} 
      \end{minipage}
  \end{center}
\end{figure}

\begin{figure}[htbp]
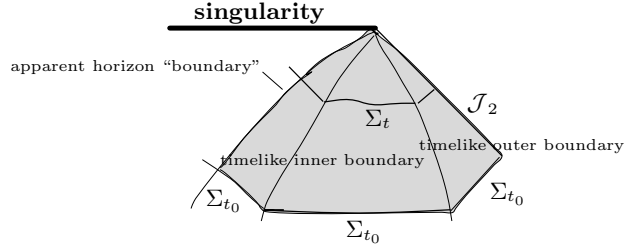

  \begin{center}
      \begin{minipage}[t]{8cm}
        \input CauCharKruskal.pstex_t
        \\
        \caption{\label{CauChar}The part of the Penrose diagram of Schwarzschild
          spacetimes which is calculated in the Cauchy characteristic
          approach. Again every point represents an $S^2$.} 
      \end{minipage}
  \end{center}
\end{figure}

We now describe how the same situation is approached in two alternative
numerical approaches, namely the approach of solving the initial boundary value
problem in physical spacetime and the so-called Cauchy
characteristic matching approach.
\\
Figure~\ref{KakKruskal} shows the Penrose diagram of the first case,
the approach by solving the initial boundary value problem in physical
spacetime as described in~\cite{AnCA95td}.
Data are given on a compact part of a spacelike initial slice, which
is covered by gridpoints, and on the boundaries.
To be representative for the whole slice, which has topology $R\times S^2$,
the compact part should have topology $I\times S^2$, where $I$ is a
closed interval.
The inner edge of that closed interval $I$ is placed somewhere
inside the event horizon, since then its treatment will not influence
the physical predictions.
For technical reasons the inner ``boundary'' is often placed near the apparent 
horizon and therefore called apparent horizon boundary, although the
apparent horizon is a null or spacelike hypersurface.
In contrast to the inner edge the outer edge of the grid is treated in
such a way that it forms a timelike boundary under time evolution.
Due to the problem of not having a consistent outer boundary treatment 
without artificial back scattering, the numerical solution will be
consistent with the Einstein equation in the hatched triangle only.
The actual size of the hatched region depends on how much gridpoints
one can afford to provide to cover the parts of the hypersurface
far away.
Assuming a consistent outer boundary treatment one would get the
shaded region.
Null infinity can never be covered, even after 
rescaling, for the spacelike surfaces used go to spacelike infinity as
sketched by the dashed lines.
\\
The approximate radiation extraction methods suggested require being
``far away'' to give a reasonable approximation for the actually
emitted radiation.
Nevertheless reading off radiation at a finite distance causes an
approximation error.
Hence the error made will not converge to zero when refining the grid.
It has therefore been suggested to combine the grid refinements with
placing the outer boundary further out.
There are two problems with this ``solution''.
Firstly, the number of required gridpoints grows much faster than in
the case of only refining and this additional growth certainly outweighs
the benefits from computational science efforts to optimise memory
footprint and execution speed per gridpoint.
Secondly, whenever we place the boundary at a different distance we
solve a different problem.
This change of the problem does not occur in the conformal method,
since the grid contains for every refinement level the conformally
completed spacetime and null infinity.
\\
The second alternative we compare with, Cauchy characteristic
matching~\cite{GoLA98sc}, is shown in figure~\ref{CauChar}.
Near the outer edge and near the apparent horizons the spacetime is
sliced by null hypersurfaces.
The inner null hypersurfaces are cut off inside the apparent horizon.
The region between the null hypersurfaces is evolved via a Cauchy
problem.
As null hypersurfaces unavoidably develop caustics for strong
data, the Cauchy slicing in the middle is necessary and must stretch out
sufficiently far.
Matching the two schemes introduces artificial boundaries and causes
stability problems~\cite{GoLA98sc}.
Although null infinity is mapped to gridpoints by a coordinate
rescaling, null infinity remains a true numerical boundary as the
equations used are singular there.

As a conclusion we find that in all approaches only part of the whole
spacetime can be calculated.
But this does not favour the other approaches discussed, because none
of them allows us to calculate a larger part of the Penrose diagram.
Even when using the data on $\Sigma_{t_0}$ to integrate backwards in
time, as done in~\cite{Hu93ns}[figure Va], the neighbourhood of the
$i^0$s cannot be obtained.
To obtain the whole Penrose diagram one would have to start with an
initial slice containing spacelike infinity.
The conformal metric at $i^0$ is in general not sufficiently smooth, 
some of the variables used in the conformal field equations are not
well-behaved at $i^0$.
Therefore the more advanced and complicated conformal techniques
of~\cite{Fr96gf} would be required.
\subsection{``Tracking radiation sufficiently far'' or ``The need for
  apparent horizons''}
\label{Vergleich}
\begin{figure}[htbp]
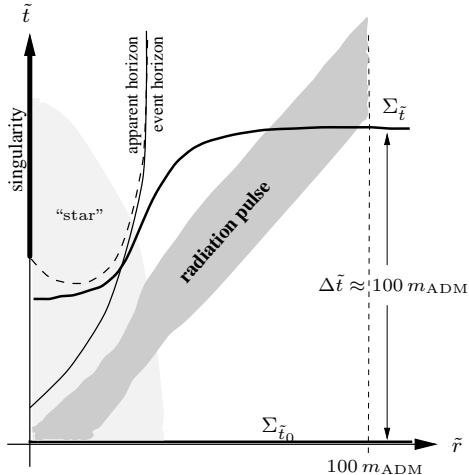

  \begin{center}
    \begin{minipage}[t]{6.0cm}
      \input PhysikalischeSlicing.pstex_t
      \\
      \caption{\label{PhysikalischeSlicing}Cauchy slices in physical
        spacetime}
    \end{minipage}
  \end{center}
\end{figure}

Probably the simplest scenario which involves gravitational radiation
and for which apparent horizon boundary conditions have been regarded
as necessary is an energy/matter concentration collapsing to a black
hole.
\\
Figure~\ref{KonformesSlicing} shows in a diagram how this scenario is
calculated in the conformal approach.
Its correctness has been justified by the model calculations discussed
in~\cite{Hu93nu,Hu96mf}.
\begin{figure}[htbp]
  \begin{center}
      \begin{minipage}[t]{6.6cm}
        \input KonformesSlicing.pstex_t
        \\
        \caption{\label{KonformesSlicing}Cauchy slices in conformal
          spacetime (= hyperboloidal slices in physical spacetime)}
      \end{minipage}
    \hspace{4em}
      \begin{minipage}[t]{7cm}
        \input KonformesPhysSlicing.pstex_t
        \\
        \caption{\label{KonformesPhysSlicing}Hyperboloidal slices in
          physical spacetime}
      \end{minipage}
  \end{center}
\end{figure}
In the situation drawn a singularity avoiding slicing of the
conformal spacetime is in no way pathologic --- null infinity,
timelike infinity, and the singularity are all at finite conformal
coordinate time.
A pulse of radiation will propagate outward with a slope of order one
and will cross \Scri{} at finite conformal coordinate time.
When this pulse crosses \Scri{} we can read off the radiation which is 
seen by an observer at infinite physical distance.
Although the physical distance between the gridpoints becomes very
large near \Scri{}, the number of gridpoints used for resolving a
signal, as indicated by the width of the pulse, stays approximately
constant.
The mathematical reason behind this feature is the invariance of the
light cone under conformal rescalings.
\\
If we translate figure~\ref{KonformesSlicing} to physical spacetime
coordinatized by physical time $\tilde t$ and by physical distance
$\tilde r$ we get figure~\ref{KonformesPhysSlicing}.
The hyperboloidal character of the hypersurfaces $\Sigma_t$ becomes
obvious, for large distances the hypersurfaces approach null
hypersurfaces.
Since the hypersurfaces are almost null far out, the finite, but far
distance at which we want to read off radiation does not significantly 
influence the integration time required.
In the example shown we would need to calculate a numerical time
evolution for $\tilde t = 35\,m_{\rm ADM}$ to be able to read off
radiation at $\tilde r = 100\,m_{\rm ADM}$.
\\
We also see that these ``physical'' coordinates do not very well
represent the causal structure of the spacetime near and inside the
event horizon --- the singularity and the event horizon look like
timelike lines.
Furthermore we notice that in the interior the slices cease to be
regular after a finite time, whereas in the exterior they exist
forever.
Therefore, if we would like to calculate the longtime behaviour as
seen by an observer outside the event horizon we must either
distort the hypersurface somewhere or cut out the interior part.
\\
Figure~\ref{PhysikalischeSlicing} shows the situation in the physical
spacetime with singularity avoiding slices, the slicing which is
believed to be the most appropriate way for approaching the problem in
physical spacetime.
It is obvious that if we want to read off radiation at $\tilde r =
100\,m_{\rm ADM}$, we need to calculate the time evolution up to
$\tilde t = 100\,m_{\rm ADM}$, and if we want to read it off at
$\tilde r = 1000\,m_{\rm ADM}$, we even need to integrate to $\tilde t
= 1000\,m_{\rm ADM}$.
Therefore, even if we are only interested in gravitational radiation but
not in the longtime behaviour of the spacetime, it is necessary to
numerically integrate for a long time, at least in the exterior.
To not hit the singularity when doing so one again either has to deal
with a significant distortion of the hypersurfaces, i.~e.\ large
gradients of the variables somewhere on the hypersurfaces, or one has
to cut out the interior of the hypersurfaces, which is the purpose of
using apparent horizon boundary conditions.
\\
From what has been said about figure~\ref{KonformesPhysSlicing}
and~\ref{PhysikalischeSlicing} one might be tempted to utilize
hyperboloidal slices in the physical spacetime for the numerical
calculation.
This would, unfortunately, introduce new problems.
They would arise from the fact that the evolution equations require 
spacelike surfaces but the hyperboloidal slices are almost null in the 
far zone.
Technically, the lapse and/or shift degenerate when a hypersurface
becomes almost null.
Since in conformal spacetime the hyperboloidal hypersurface is a
normal Cauchy surface even at null infinity and beyond, this problem
does not arise in the conformal approach.
%
%
%
%

%
%
\section{Conclusion}
Starting from the conformal field equations we have derived a
symmetric hyperbolic first order system determining the time
evolution.
The system obtained is in a form which allows us to apply standard
techniques to discretise in the interior as well as on the grid boundaries.
The stability of the resulting schemes is then guaranteed by theorems
in the literature.
\\
We have seen that an application of only these techniques allows us in
principle to calculate the complete future of the initial data for
many interesting scenarios, e.~g.\ N black holes.
\section*{Acknowledgement}
I would like to thank all the visitors and members of the Albert
Einstein institute who where available for numerous discussions
concerning the conformal field equations and their numerical
implementation.
\\ 
In particular I would like to mention H.~Friedrich, B.~Schmidt, and
J.~Frauendiener for many fruitful discussions.
%
%
%
%

\bibliography{biblio}
\bibliographystyle{prsty}

\end{document}